\documentclass[conference]{IEEEtran}
\IEEEoverridecommandlockouts
\usepackage{cite}
\usepackage[super]{nth}
\usepackage{amsmath,amssymb,amsfonts}
\usepackage{algorithmic}
\usepackage{makecell}
\usepackage{url}
\usepackage{graphicx}
\usepackage{textcomp}
\usepackage{xcolor}
\usepackage{booktabs}
\usepackage{multirow}
\usepackage[textsize=tiny]{todonotes}
\usepackage{hyperref}

\def\BibTeX{{\rm B\kern-.05em{\sc i\kern-.025em b}\kern-.08em
    T\kern-.1667em\lower.7ex\hbox{E}\kern-.125emX}}
    
\begin{document}

\title{%
    Anchoring the Unknown: Open-Set Model Attribution via Proxy-Anchor Learning%
    \thanks{
        This work was in part supported by a grant of the Ministry of Research, Innovation and Digitization, CCCDI - UEFISCDI, project number PN-IV-P7-7.1-PTE-2024-0600, within PNCDI IV.
        This work was also partially supported by the European Union – NextGenerationEU, through the National Recovery and Resilience Plan (PNRR), Component 9, Investment 4, under project SENSE, THINK @ POLITEHNICA BUCUREȘTI (SENTHIPoli) No. RUE 6.PI/I4/C9.
        The content of this material does not necessarily represent the official position of the European Union or the Government of Romania.
    }%
}




\author{
    \IEEEauthorblockN{
        Cristian-Teodor Neamtu\textsuperscript{1},
        Serban Mihalache\textsuperscript{1},
        Stefan Smeu\textsuperscript{2},
        Dan Oneata\textsuperscript{1, 2},
        Horia Cucu\textsuperscript{1},
        Dragos Burileanu\textsuperscript{1}
    }
    \IEEEauthorblockA{\textsuperscript{1}Speech and Dialogue Research Laboratory (SpeeD), POLITEHNICA Bucharest \,\,\,\, \textsuperscript{2}Bitdefender}
    \IEEEauthorblockA{
        \{cristian.neamtu, serban.mihalache, dan\_theodor.oneata, horia.cucu, dragos.burileanu\}@upb.ro, 
        ssmeu@bitdefender.com
    }
}

\setlength{\marginparwidth}{1.65cm}

\definecolor{myblue}{HTML}{1f78b4}
\definecolor{mybluelight}{HTML}{a6cee3}

\definecolor{mygreen}{HTML}{fb9a99}
\definecolor{mygreenlight}{HTML}{33a02c}

\definecolor{mypurple}{HTML}{6a3d9a}
\definecolor{mypurplelight}{HTML}{cab2d6}

\definecolor{myred}{HTML}{fdbf6f}
\definecolor{myredlight}{HTML}{e31a1c}

\definecolor{myolive}{HTML}{ada136}
\definecolor{myolivelight}{HTML}{7ead36}

\newcommand{\dan}[1]{\textcolor{mypurple}{#1}}
\newcommand{\horia}[1]{\textcolor{myred}{#1}}
\newcommand{\stefan}[1]{\textcolor{myolive}{#1}}
\newcommand{\cristi}[1]{\textcolor{red}{#1}}

\newcommand{\tododan}[1]{\todo[color=mypurplelight]{Dan: #1}}
\newcommand{\todohoria}[1]{\todo[color=myredlight]{Horia: #1}}

\newcommand{\ab}{\mathbf{a}}
\newcommand{\eb}{\mathbf{e}}
\newcommand{\pb}{\mathbf{p}}

\maketitle

\begin{abstract}
    The proliferation of text-to-speech (TTS) systems capable of generating realistic synthetic speech poses growing challenges for audio forensics. While binary deepfake detection has received considerable attention, source tracing (i.e., identifying which TTS system produced a given audio sample) remains underexplored, particularly in open-set scenarios where unknown systems may be encountered. We propose a metric learning framework based on the Proxy-Anchor loss function that operates on Wav2Vec2-BERT embeddings to learn a discriminative embedding space for TTS source attribution and out-of-distribution (OOD) detection of unseen systems. We evaluate it on the MLAAD v9 dataset spanning 140 TTS systems across 51 languages, and introduce an architecture merging strategy that groups TTS system versions into unified classes, reducing inter-class confusion. Our system achieves 99.76\% accuracy on 110 in-distribution classes and a False Positive Rate (FPR@95) as low as 2.04\% for OOD detection. Also, for a fair comparison against the current state of the art, we further evaluate it on the MLAAD v5 official dataset splits, improving the OOD accuracy by almost doubling it. These results demonstrate that Proxy-Anchor metric learning, combined with architecture-aware class design and post-hoc OOD scoring, provides an effective framework for forensic TTS source tracing in both closed-set and open-set settings.
\end{abstract}


\section{Introduction}

Recent advances in text-to-speech (TTS) and voice conversion (VC) technology have made it possible to generate synthetic speech that is increasingly difficult to distinguish from human recordings \cite{wang_neural_2023, tan_naturalspeech_2022, casanova_yourtts_2023}. With the rapid growth of publicly available generative systems, the potential for misuse in fraud, disinformation, and identity impersonation has become a pressing concern. While substantial research has been devoted to audio deepfake detection as a binary classification task \cite{liu_asvspoof_2023} (i.e., determining whether a given audio sample is real or synthetic), this alone is insufficient for forensic analysis. When a synthetic sample is identified, investigators often need to determine which system produced it, a task known as source tracing or attribution.

Source tracing has received growing attention in recent years, with approaches varying in granularity and problem formulation. Klein et al. \cite{klein_source_2024-1} proposed a two-stage pipeline that first detects spoofed audio and then classifies spoofing attributes such as the acoustic model and vocoder, evaluating on both ASVspoof 2019 \cite{wang_asvspoof_2020} and MLAAD \cite{muller_mlaad_2026}. Their subsequent work \cite{klein_open-set_2025} extended this to the open-set scenario, introducing softmax energy as a novel adaptation of the energy score for out-of-distribution (OOD) detection, achieving a False Positive Rate (FPR@95) of 8.3\% with augmentation strategies. Doan et al. \cite{doan_vib-based_2025} introduced a real pre-emphasis method based on the Variational Information Bottleneck architecture combined with supervised contrastive learning, improving generalization on MLAAD v5 by 10\% over baselines. Falez et al. \cite{falez_audio_2025} addressed source tracing through multi-attribute open-set identification and verification protocols, classifying either the full generative system or its individual components, and emphasized the need for a standardized ontology in this domain.

Several recent works have explored metric learning for source tracing, drawing on techniques from speaker recognition. Koutsianos et al. \cite{koutsianos_synthetic_2025} compared classification-based and metric learning approaches using ResNet and self-supervised learning (SSL) backbones on MLAAD v5, demonstrating that ResNet with metric learning can match or exceed SSL-based systems. Negroni et al. \cite{negroni_source_2025} introduced the source verification task, inspired by speaker verification, where the goal is to determine whether a test track was produced by the same model as a set of reference signals, evaluating robustness across speaker diversity, language mismatch, and post-processing conditions. Stan et al. \cite{stan_tada_2025} proposed TADA, a training-free approach based entirely on $k$-Nearest Neighbors over pre-trained SSL embeddings, achieving an F1-score of 0.93 for attribution and 0.84 for OOD detection. Xuan et al. \cite{xuan_multilingual_2025} established the first benchmark for multilingual source tracing, investigating both DSP- and SSL-based features and evaluating cross-lingual generalization to unseen languages and speakers.

Despite this progress, several limitations persist. Most closed-set approaches assume a fixed set of known source systems, limiting applicability as new TTS systems emerge. Open-set methods have typically been evaluated with a small number of held-out systems. Furthermore, many TTS systems exist as versions of the same architecture (differing only in model size, training data, or version) leading to high inter-class confusion when treated as independent categories.

Our contributions are: \textbf{(1)} an application of the Proxy-Anchor loss function to audio deepfake source tracing, \textbf{(2)} a new strategy for grouping model architectures, \textbf{(3)} experiments on MLAAD v9 achieving 99.76\% closed-set accuracy and 2.04\% FPR@95 for OOD detection, and \textbf{(4)} demonstrating superior performance compared to the current state of the art.

\section{Methodology} \label{Proposed System Architecture}

We formulate the source tracing task as a non-binary classification problem.
Specifically, we train a model that maps a given audio to one of $K$ classes, corresponding to the speech synthesis system used for generating the input audio.
Our approach is based on self-supervised audio representations and learns a prototype for each class, which is used for prediction.
Given the continuous emergence of new synthesis systems, we also investigate approaches that identify those samples that do not belong to any of the known classes, i.e., the task of out-of-distribution (OOD) detection.

\textbf{Audio representations.}
To encode the input audio, we use the Wav2Vec2-BERT model \cite{communication_seamless_2023}.
This self-supervised model is trained on 4.5M hours of unlabeled data covering over 143 languages.
Its architecture is based on Conformer layers, which combine Convolutional Neural Networks and
Transformers to model both local and global dependencies.
We extract embeddings from the fourth layer (out of 24), as earlier layers have been found to capture more discriminative features for deepfake detection \cite{kheir_comprehensive_2025,pascu2025detecting}.
The resulting representations are frame-level features, which we aggregate into a fixed 1024-dimensional utterance-level embedding using temporal mean pooling.
Prior work has also shown the benefits of other pre-trained audio models
(Wav2Vec2 \cite{wang22odyssey, martin-donas_exploring_2024,pascu24interspeech}, 
WavLM \cite{combei2024wavlm}, 
HuBERT \cite{kheir_comprehensive_2025}, 
Whisper \cite{kawa2023improved}). We also evaluated WavLM-Large \cite{Chen_2022} and HuBERT-Large \cite{hsu2021hubertselfsupervisedspeechrepresentation}. Although they yielded similar results, Wav2Vec2-BERT works slightly better on the in-distribution problem, thus we selected it for all experiments.

\textbf{Model.}
The audio representations $\ab$ are fed into a trainable projection head $\phi$, which is implemented as a linear layer that maps the  1024-dimensional embedding space to a 1024-dimensional output space.
Each class is modeled by a  learnable prototype embedding $\mathbf{p}$, which is optimized jointly with the projection head using a contrastive loss.
Given an input sample, the model computes a class score  for class $k$ as the cosine similarity between the projected embedding and the corresponding proxy:
    $s_k =  \langle \phi(\ab), \pb_k \rangle$.
The predicted class is the one whose prototype yields the highest similarity score.

%
%

\textbf{Loss function.}
To optimize the model's parameters (the linear layer and the $K$ proxies), we minimize the  Proxy-Anchor loss function \cite{kim_proxy_2020}.
The loss treats each proxy as an anchor and it pushes embeddings of other classes away from the proxy.
By utilizing proxies, the model learns a holistic representation of each generative architecture.
This global view prevents the network from overfitting to specific samples and ensures that the embedding space is tightly clustered, which is critical for detecting OOD attacks later.
After the training phase, we obtain the optimized projection head and the final set of class-representative proxies.
As previously outlined, our inference framework operates in two distinct stages: OOD detection and in-distribution (ID) attribution.

\textbf{OOD detection.}
To tell whether a query sample belongs to a system seen at training time or is an OOD sample, we experiment with three scoring functions:
\begin{itemize}
    \item \textit{Softmax energy} \cite{klein_open-set_2025} 
    normalizes the similarities $s_k$ using the softmax function $\sigma$ and aggregates them using log-sum-exp, as shown in (\ref{eq:softmax-energy}).
    \begin{equation} \label{eq:softmax-energy}
        - \log \sum_{k=1}^{K} \exp {\sigma(s_k)}
    \end{equation}
    \item \textit{Shannon entropy} %
    measures the class distribution-related uncertainty of the softmax-normalized similarities, given in (\ref{entropy}).
    \begin{equation} \label{entropy}
         -\sum_{i=1}^{K} \sigma\!\left(s_k\right) \cdot \log\, \sigma\!\left(s_k\right)
    \end{equation}
    \item \textit{Maximum proxy distance} 
    measures the cosine distance to the most similar proxy, as per (\ref{max_proxy-similarity}). 
    \begin{equation}\label{max_proxy-similarity}
        1 - \max_{k} \, \langle\phi(\ab), \mathbf{p}_k\rangle
    \end{equation}  
\end{itemize}

The larger the score, the more likely it is that the sample is OOD.
To obtain a binary decision, we classify a sample as OOD if the score is greater than a threshold $\tau$.
The threshold is set such that the True Positive Rate (TPR) is 95\%
on a held-out OOD calibration set.
%
%
%
%

\textbf{Inference.}
Evaluation is performed hierarchically.
For an embedding extracted by the backend pre-trained model, we compute its uncertainty score and compare it against the pre-calibrated threshold $\tau$.
If the uncertainty exceeds the threshold, the sample is immediately flagged as OOD, signifying an unknown generative architecture.
Otherwise, the sample satisfies the ID condition and it proceeds to the attribution stage, where the system classifies the sample by identifying the nearest class proxy in the embedding space.
Formally, the predicted label $\hat{y}$ is determined by maximizing the cosine similarity between the embedding and the learned proxies.

\section{Experimental Setup}

\subsection{Dataset and Partitioning}

Our experiments are performed on MLAAD v9 (Multi-Language Audio Anti-Spoofing Dataset) \cite{muller_mlaad_2026}.
This dataset comprises 678.3 hours of synthetic speech generated by 140 different TTS system across 51 languages.
%
We consider two experimental settings:
\begin{itemize}
\item \textit{Experiment 1}: Each of the 140 TTS systems is treated as a distinct class.
This setting evaluates the model's ability to distinguish between closely related versions of the same generator (e.g., Llasa-1B and Llasa-3B).
\item \textit{Experiment 2}:
Systems sharing the same underlying architecture are merged into a single class.
The merging is done manually based on the model versions and results in 
130 classes.
\end{itemize}

In both experiments, 20 systems are reserved for OOD evaluation (10 for calibration and 10 for testing), while the remaining systems are used as in-distribution (ID) classes, as shown in Table~\ref{tab:dataset_splits}.
The ID set is further partitioned into three distinct subsets (training, validation, testing), using a stratified random split with a 14:3:3 ratio.
To prevent class imbalances,
we cap the number of utterances per TTS system at 2\,000 (the median value across systems).

\begin{table}
    \centering
    \caption{Dataset partitioning for the two experimental protocols.}
    \label{tab:dataset_splits}
    \renewcommand{\arraystretch}{1.2}
    \begin{tabular}{l c c}
        \toprule
        & \multicolumn{2}{c}{\textbf{\# Systems}} \\ 
        \cmidrule(lr){2-3} 
       \textbf{Subset} &  \textit{Experiment 1} & \textit{Experiment 2} \\
        \midrule
        \textbf{ID}  & 120 & 110 \\
        \textbf{OOD Calibration} & 10 & 10 \\
        \textbf{OOD Test} & 10 & 10 \\
        \midrule
        \textbf{Total} & \textbf{140} & \textbf{130}  \\
        \bottomrule
    \end{tabular}
\end{table}



\subsection{Baseline Methods}

We compare our approach against two established traditional classifiers:
logistic regression and $k$-Nearest Neighbors ($k$-NN).
To ensure a fair comparison, both baselines use the same Wav2Vec2-BERT embeddings as input features.
For logistic regression, OOD detection is performed using the entropy-based scoring.
For $k$-NN, inspired by the results obtained in \cite{stan_tada_2025}, we set $k=21$ and perform OOD detection based on cosine distance;
the OOD threshold is calibrated using the mean distance to the 21 nearest neighbors.




\subsection{Implementation Details}

The feature extractor is the 
 {\fontfamily{qcr}\selectfont facebook/w2v-bert-2.0} model from HuggingFace.
 The trainable projection head consists of a single linear layer that maps the 1024-dimensional input features to a 1024-dimensional metric space.
 We apply L2 normalization to the output embeddings, constraining them to lie on the unit hypersphere.

The model is implemented in PyTorch and trained on six NVIDIA Tesla T4 GPUs (16 GB VRAM each)%
We trained the projection head using AdamW with an initial learning rate of $10^{-3}$ and a weight decay of $10^{-4}$.
For the Proxy-Anchor loss function, we chose standard hyperparameters with a margin $\delta=0.1$ and a scaling factor $\alpha=32$.
The batch size was set to 256.
The model was trained for 100 epochs, 
selecting the checkpoint with the highest validation accuracy.

Code and corresponding data protocols are available at \texttt{https://github.com/neamtucristian26/panda}.


\subsection{Evaluation Metrics}

To evaluate performance, we used different metrics for the two stages of our pipeline:
\begin{enumerate}
    \item \textit{OOD detection}: To measure the separation between ID and OOD samples, we report the Area Under the Receiver Operating Characteristic Curve (AUROC), precision, F1-score, and False Positive Rate at 95\% True Positive Rate (FPR@95).
    \item \textit{Model attribution}: For samples correctly identified as ID, we report the classification accuracy to assess the model's capability to attribute the correct generator.
\end{enumerate}

\section{Evaluation and Results}

\subsection{Closed-set Attribution}

Table \ref{tab:classification} reports the ID classification accuracy for all methods under both experimental protocols. In Experiment~1 (120 individual TTS systems), our Proxy-Anchor model achieves 98.23\% accuracy, comparable to Logistic Regression (98.16\%) and substantially outperforming $k$-NN (92.58\%). In Experiment~2 (110 merged architectures), all methods showed improved performance, with our approach reaching 99.76\%. The consistent gains under merging (+1.53\% for Proxy-Anchor) confirm that grouping model versions by their underlying architecture reduces inter-class ambiguity. The near-identical performance of Proxy-Anchor and Logistic Regression on ID classification suggests that the representations extracted by the frozen Wav2Vec2-BERT model are already highly linearly separable; the advantage of metric learning manifests primarily in the OOD scenario. 

\begin{table}[htbp]
    \centering
    \caption{ID attribution accuracy [\%].}
    \label{tab:classification}
    \renewcommand{\arraystretch}{1.15}
    \begin{tabular}{l c c}
        \toprule
        \textbf{Method} & \textbf{Exp.~1} & \textbf{Exp.~2} \\
        & (120 classes) & (110 classes) \\
        \midrule
        $k$-NN ($k$=21) \cite{stan_tada_2025} & 92.58 & 95.15 \\
        Logistic Regression                   & 98.16 & 99.59 \\
        \textbf{Proxy-Anchor (Ours)}          & \bf 98.23 & \textbf{99.76} \\
        \bottomrule
    \end{tabular}
\end{table}

\subsection{OOD Detection}

Table \ref{tab:ood_detail} presents the OOD detection results for all three scoring functions in the Proxy-Anchor framework. We select the best method per experiment based on the lowest FPR@95, as minimizing false alarms is paramount in forensic scenarios.

\begin{table}[b]
    \centering
    \caption{OOD detection performance for Proxy-Anchor \\ under both experimental protocols [\%]. \\ Best FPR@95 per experiment is \textbf{bolded}.}
    \label{tab:ood_detail}
    \renewcommand{\arraystretch}{1.15}
    \begin{tabular}{l l r r r r}
        \toprule
        \textbf{Exp.} & \textbf{Scoring Method} & \textbf{AUROC} & \textbf{FPR@95} & \textbf{Prec.} & \textbf{F1} \\
        \midrule
        \multirow{4}{*}{1}
        & Softmax energy  & 98.32 & 10.74 & 84.57 & 90.12 \\
        & Entropy         & 97.98 & \textbf{9.59} & 85.63 & 89.47 \\
        & Max proxy dist.  & 98.54 & 11.29 & 83.99 & 90.05 \\
        \midrule
        \multirow{4}{*}{2}
        & Softmax energy & 99.14 & 3.15 & 95.82 & 94.69\\
        & Entropy        & 99.08 & 3.53 & 95.37 & 94.77 \\
        & Max proxy dist. & 99.35 & \textbf{2.04} & 97.18 & 94.10 \\
        \bottomrule 
    \end{tabular}
\end{table}

\begin{figure*}[htb]
    \centering
    \includegraphics[width=0.95\textwidth]{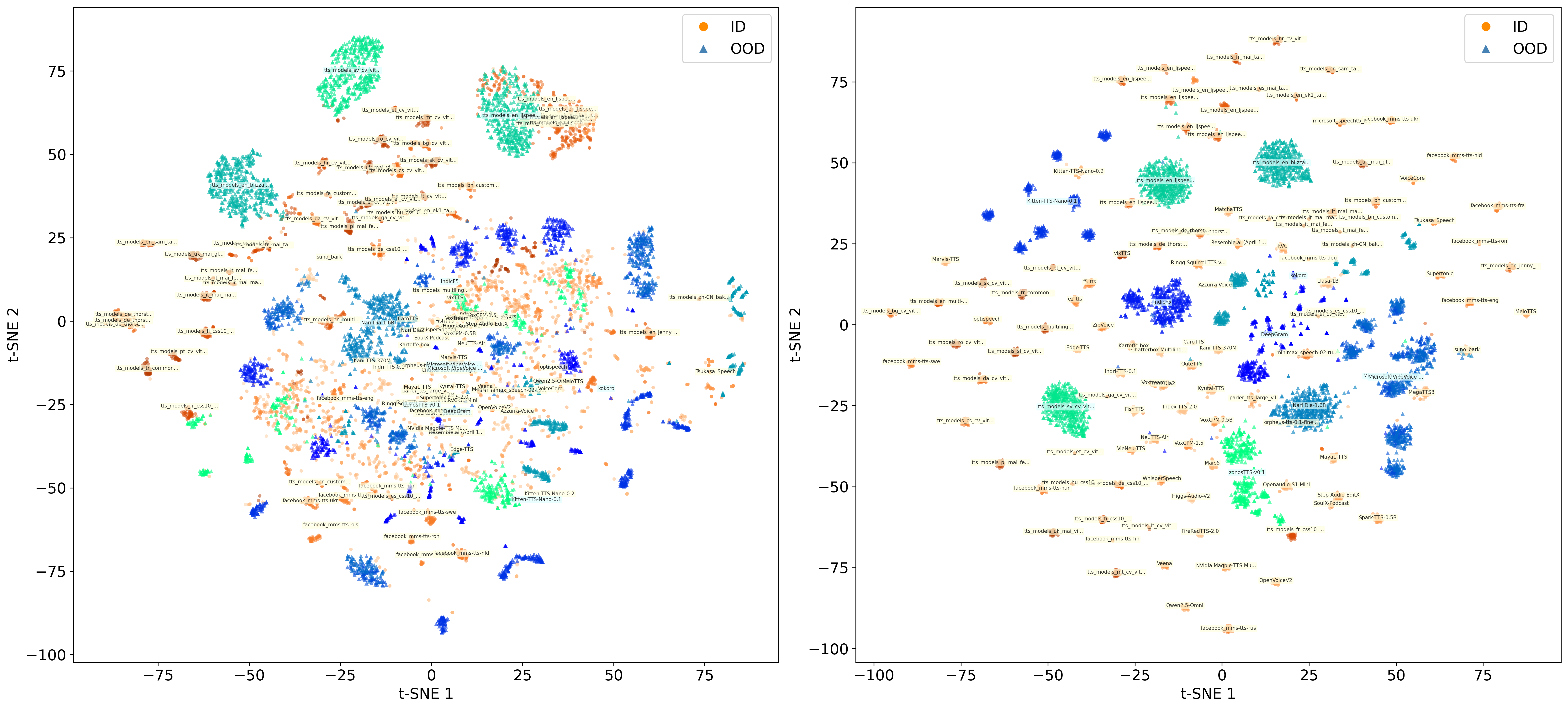}
    \caption{t-distributed Stochastic Neighbor Embedding (t-SNE) visualization of embedding spaces. \textbf{Left}: Raw Wav2Vec2-BERT embeddings showing overlap between in-distribution samples (warm-colored circles) and OOD samples (cold-colored triangles). \textbf{Right}: Proxy-Anchor projected embeddings demonstrating enhanced separation with in-distribution clusters tightly grouped and OOD samples pushed away, validating the 99.4\% AUROC and 2.0\% FPR@95 achieved by the best proxy-similarity scoring method.}
    \label{fig:id_ood_embeddings}
\end{figure*}

In Experiment~1, Shannon entropy achieves the best FPR@95 of 9.59\% with an AUROC of 97.98\%. The entropy score captures prediction uncertainty effectively when the class count is large (120 systems) since ID samples tend to produce sharply peaked softmax distributions.

In Experiment~2, the maximum proxy distance leads to top performance with an FPR@95 of only 2.04\% and an AUROC of 99.35\%. This scoring function directly exploits the structure of the Proxy-Anchor embedding space. After merging, the proxies are more tightly clustered around distinct architectures, yielding a more pronounced separation between ID and OOD samples in similarity space.


Fig. \ref{fig:id_ood_embeddings} provides visual confirmation of these quantitative results, showing that the Proxy-Anchor projection substantially enhances the geometric separation between ID and OOD samples compared to raw Wav2Vec2-BERT embeddings.

\subsection{Cross-Method Comparison}

A unified comparison of all evaluated systems, using the best-performing OOD metric for each method, is provided in Table \ref{tab:cross_method}. Across both experiments, Proxy-Anchor achieves the lowest FPR@95 while maintaining competitive ID classification accuracy. The FPR@95 advantage over Logistic Regression is 4.1\% in Experiment~1 and 14.5\% in Experiment~2. The $k$-NN baseline exhibits poor OOD discrimination (FPR@95 of 54.63\% and 66.90\%), indicating that raw embedding distances without learned projections are insufficient for reliable open-set detection.

These results demonstrate that the metric learning objective provides a dual benefit: it preserves the discriminative power of the self-supervised features for ID classification while simultaneously structuring the embedding space in a way that facilitates uncertainty-based OOD detection.

\begin{table}[htbp]
    \centering
    \caption{Cross-method comparison using each system's best OOD scoring method [\%]. Best FPR@95 per experiment is \textbf{bolded}.}
    \label{tab:cross_method}
    \renewcommand{\arraystretch}{1.15}
    \begin{tabular}{l c r r}
        \toprule
        \textbf{Method} & \textbf{Exp.} & \textbf{AUROC} & \textbf{FPR@95} \\
        \midrule
        k-NN ($k$=21) \cite{stan_tada_2025} & 1 & 82.11 & 54.63     \\
        Logistic Regression                 & 1 & 97.02 & 13.73 \\
        \textbf{Proxy-Anchor (Ours)}        & 1 & 97.98 & \textbf{9.59} \\
        \midrule
        k-NN ($k$=21) \cite{stan_tada_2025} & 2 & 78.48 & 66.90 \\
        Logistic Regression                 & 2 & 97.13 & 16.51 \\
        \textbf{Proxy-Anchor (Ours)}        & 2 & 99.35 & \textbf{2.04} \\
        \bottomrule
    \end{tabular}
\end{table}

\subsection{Comparison with the State of the Art}

While our initial experiments were conducted on MLAAD v9, we report results on MLAAD v5 to ensure a fair and direct comparison with prior work, as existing state-of-the-art methods are evaluated on this version of the dataset.

As shown in Table~\ref{tab:sota_comparison}, our method consistently outperforms both baselines across all reported metrics.
In closed-set attribution, we achieve an ID accuracy of 98.57\%, improving upon both Kulkarni et al. (95.61\%) and Klein et al. (95.80\%). 

\begin{table}[b]
    \centering
    \caption{Comparison with state-of-the-art source tracing algorithms on MLAAD v5 official splits \cite{UsingMLAADforSourceTracing}. We report accuracy [\%] for closed-set attribution and accuracy [\%] and FPR@95 [\%] for OOD detection.}
    \label{tab:sota_comparison}
    \renewcommand{\arraystretch}{1.15}
    \begin{tabular}{lccc}
        \toprule
         & \textbf{In-distribution} & \multicolumn{2}{c}{\textbf{Out-of-distribution}} \\
        \cmidrule(lr){2-2} \cmidrule(lr){3-4}
        \textbf{Method} & \textbf{Acc.} & \textbf{Acc.} & \textbf{FPR@95}\\
        \midrule
        Kulkarni et al. \cite{kulkarni_unveiling_2025} & 95.61 & 44.82 & -- \\
        Klein et al. \cite{klein_open-set_2025} & 95.80 & -- & 8.30\\
        \textbf{Ours} & \textbf{98.57} & \textbf{89.20} & \textbf{3.36} \\
        \bottomrule
    \end{tabular}
\end{table}

The most striking gains, however, are in OOD detection.
Our system achieves an OOD accuracy of 89.20\%, nearly doubling the 44.82\% reported by Kulkarni et al. (Klein et al. do not report this metric).
Similarly, our FPR@95 of 3.36\% represents a 60\% relative reduction compared to the 8.30\% of Klein et al., indicating that our system misclassifies far fewer known generators as unknown ones.
Notably, ours is the only method that demonstrates strong performance on both closed-set attribution and open-set OOD detection simultaneously, suggesting that the two objectives are complementary rather than competing in our framework.

\section{Conclusion}

We addressed the source tracing task of TTS systems, with a particular focus on an open-set context.
Our approach combines Proxy-Anchor metric learning with multiple strategies for out-of-distribution (OOD) scoring.
On MLAAD v5, our system achieves 98.57\% closed-set accuracy and 3.36\% FPR@95, substantially outperforming recent baselines:
we improve OOD accuracy by almost doubling it compared to Kulkarni et al. \cite{kulkarni_unveiling_2025} and
reduce FPR@95 by 60\% relative to Klein et al. \cite{klein_open-set_2025}.
When scaled to the MLAAD v9 benchmark with 110 merged architectures, performance further improves to 99.76\% closed-set accuracy and 2.04\% FPR@95, demonstrating that our approach scales favorably with the number of TTS systems.

A key finding is that Proxy-Anchor metric learning produces embeddings where closed-set attribution and open-set rejection are complementary: well-clustered class representations naturally yield reliable uncertainty estimates for unknown generators.
Architecture-level merging further reinforces this effect, reducing FPR@95 by a factor of 4.7 compared to non-merged configurations.

From a forensics perspective, our system enables provenance tracing to architecture families across 51 languages and 140 systems, with very low false alarm rates, suitable for practical deployment.
Future work will address robustness to in-the-wild conditions and adversarial attacks, and few-shot adaptation to newly emerging TTS architectures.



\bibliography{references, refs-dan}{}
\bibliographystyle{ieeetr}

\end{document}